\documentclass[aps,pre,twocolumn]{revtex4-1}
\usepackage{graphicx}
\usepackage{amsmath}
\usepackage{bm}
\usepackage{amssymb}
\usepackage{booktabs}
\usepackage{float}

\usepackage{hyperref}
\hypersetup{pdftitle={Universality of spectral fluctuations in open quantum chaotic systems},
	pdfauthor={Jisha C, Ravi Prakash},
	colorlinks=true,
	citecolor=blue,
	urlcolor=blue}

\newcommand{\sG}{\text{sG}}
\newcommand{\G}{\text{G}}
\newcommand{\sdG}{\text{sdG}}

\begin{document}
\title{Universality of spectral fluctuations in open quantum chaotic systems}
\author{Jisha C} \email{jisha.2021rph03@mnnit.ac.in}
\author{Ravi Prakash}\email{ravi.prakash@mnnit.ac.in}
\affiliation{Department of Physics, Motilal Nehru National Institute of Technology Allahabad, Prayagraj -- 211004, India}

\begin{abstract}
	Quantum chaotic systems with one-dimensional spectra follow spectral correlations of orthogonal (OE), unitary (UE), or symplectic ensembles (SE) of random matrices depending on their invariance under time reversal and rotation. In this letter, we study the non-Hermitian and non-unitary ensembles based on the symmetry of matrix elements, \textit{viz.} ensemble of complex symmetric, complex asymmetric (Ginibre), and self-dual matrices of complex quaternions. The eigenvalues for these ensembles lie in the two-dimensional plane. We show that the fluctuation statistics of these ensembles are universal and quantum chaotic systems belonging to OE, UE, and SE in the presence of a dissipative environment show similar spectral fluctuations.
	The short-range correlations are studied using spacing ratio and spacing distribution. For long-range correlations, unfolding at a non-local scale is crucial. We describe a generic method to unfold the two-dimensional spectra with non-uniform density and evaluate correlations using number variance. We find that both short-range and long-range correlations are universal. We verify our results with the quantum kicked top in a dissipative environment that can be tuned to exhibit symmetries of OE, UE, and SE in its conservative limit.
\end{abstract}

\maketitle

\section{Introduction}
Statistical properties of chaotic quantum systems have been studied using Random Matrix Theory (RMT) for a long time \cite{rmp-apandey, Porter-1965, Wigner-1967, mlmehta, oxford, haakebook}. Quantum chaotic systems show similar spectral correlation as that of orthogonal (OE), unitary (UE), and symplectic (SE) ensembles, based on their invariance under time reversal and rotation.

Motivated by the symmetry of matrix elements in OE, UE, and SE, we shall study non-Hermitian and non-unitary matrices, \textit{viz.} ensemble of complex-symmetric matrices, complex asymmetric (Ginibre) matrices, and self-dual matrices of complex quaternions. We shall show that quantum chaotic systems belonging to OE, UE, and SE show spectral correlations as that of above mentioned non-Hermitian ensembles in the presence of a dissipative environment.


RMT came into existence to study the statistical properties of excited states of heavy nuclei. It is conjectured by Bohigas-Giannoni-Schmit, also known as the BGS conjecture \cite{bgs,bgs-2}, that the spectral properties of random matrices following certain symmetry requirements are same as that of quantum chaotic systems consistent with those symmetries. Nowadays, RMT is used as a benchmark to test the chaoticity of unknown systems. RMT also found applications in several other fields such as in mathematics: distribution of zeros of Riemann zeta function \cite{Keating2000}, distribution of prime numbers \cite{Firk-2009}, in finance: fluctuations and correlations in stock market \cite{Laloux-2000}, time series analysis \cite{Fossion-2013, Vinayak-2010}, in biological systems: neural systems \cite{Sompolinsky-1988}, in networking: Google network \cite{Georgeot-2010}, etc.

The one-dimensional spectrum, such as eigen-energies of chaotic Hamiltonian or eigen-angles of the unitary Floquet operator for time-periodic Hamiltonian over a period are studied using Gaussian and circular ensembles of random matrices respectively. Random matrices representing the statistical description of the Hamiltonian for conservative systems are categorized into three types based on their time reversal invariance (TRI) and rotational invariance (RI). These are known as Gaussian Orthogonal Ensemble (GOE), Gaussian Unitary Ensemble (GUE), and Gaussian Symplectic Ensemble (GSE) \cite{mlmehta, oxford, rmp-apandey, stockmann, haakebook}. These ensembles consist of real symmetric, complex hermitian, and self-dual matrices of real quaternions. GUE is applicable for systems without any TRI and RI. For systems with TRI and RI, GOE is applicable. For spin systems with TRI preserved but RI broken, GOE and GSE are applicable for integer-spin and half-integer spin systems respectively. Systems with time-periodic chaotic Hamiltonian are better understood in terms of the Floquet operator defined over a period. The statistical features of eigen-angles for such operators are the same as that of Circular ensembles of unitary random matrices. Circular orthogonal ensemble (COE), circular unitary ensemble (CUE), and circular symplectic ensemble (CSE) have the same symmetry and universality as that of GOE, GUE, and GSE respectively \cite{mlmehta, oxford, stockmann, haakebook}.

The applicability of random matrices in open quantum systems appeared after J. Ginibre introduced the concept of complex random matrices without any hermiticity or unitarity condition \cite{ginibre}. Ensemble of Ginibre matrices have their eigenvalues distributed in the two-dimensional complex plane. The imaginary part of complex eigenvalues represents the dissipation of energy or collapse of the wave function with time \cite{qdissipation, qdissipation2, haakebook, d_braun, Akemann-2019, mlmehta}. Non-Hermitian ensembles attract applications in many other systems such as disordered systems, fractional quantum Hall effect, biological and neural networks \cite{May-1972, Sompolinsky-1988, Sommers-1988, Francesco-1994}, etc.

This letter deals with the following two problems.

(a) We study the non-Hermitian and non-unitary random matrices categorized based on the symmetry of matrix elements, \textit{viz.} complex-symmetric random matrices or symmetric Ginibre ensemble (symm-GinE), complex matrices without any symmetry (Ginibre ensemble, abbreviated as GinE), and self-dual matrices of complex quaternions \textit{i.e.}, self-dual Ginibre ensemble (self-dual-GinE). We shall show that symm-GinE, GinE, and self-dual-GinE show similar spectral fluctuations as obtained in the presence of dissipation in quantum chaotic systems with TRI and RI symmetries corresponding to orthogonal, unitary, and symplectic ensembles respectively. These ensembles have also found applications in quantum scattering, one-dimensional plasma, open systems, etc \cite{Fyodorov-2003, Hastings-2001, Hamazaki-2020, Markum-1999, Markum-1999-2, Sommers-1999}.

(b) We resolve a long-standing problem of unfolding the spectra in the complex plane and evaluate the number variance. Unfolding refers to the scaling of eigenvalues such that the average spectral density becomes constant without affecting the fluctuations. The short-range fluctuations are usually measured using nearest neighbor spacing distribution (nnsd) whereas number variance is useful to study long-range correlations. The unfolding procedure in one-dimensional spectra is unique but its extension to two-dimensional spectra can have multiple solutions and leads to several ambiguities \cite{Dusa-2022,Prasad-2022, Lucas-2020, Garcia-2023}. For locally isotropic spectra, there exists procedures to locally unfold the spectrum and measure short-range statistics such as nnsd \cite{Akemann-2019}, but a generic method to unfold the spectra globally and measure the long-range statistics such as number variance is not discussed in detail to the best of our knowledge. See also ref. \cite{Markum-1999, Markum-1999-2} which deals with non-local unfolding in a weakly non-Hermitian system and ref. \cite{Ndumbe-2020} where the spectrum is already uniform in one dimension. There exist universalities specific to long-range correlations in one-dimensional spectra such as suppression of fluctuations of large wavelengths found through the saturation of number variance \cite{Ravi-PRE-2016, Ravi-PRE-2021}. We believe that these unfolding methods will also be useful to investigate such universalities in open quantum systems. 

We develop the mechanism to unfold the spectra at a non-local scale and also evaluate the number variance. We also discuss the challenges associated with the unfolding of two-dimensional spectra. The universality of spectral correlations in symm-GinE, GinE, and self-dual-GinE for both short-range and long-range correlations is verified using nnsd and number variance respectively. We also numerically find a simple expression for number variance for self-dual-GinE.

We verify all our findings on a prototype model of chaos known as quantum kicked top. It shows characteristics of COE, CUE, and CSE for different settings of its parameters. We modify the kicked top and include a dissipative term to validate our results.

\section{Matrix ensembles and their eigenvalues}
The elements of matrices in symm-GinE, GinE, and self-dual-GinE, denoted by $M_\sG, M_\G$ and $M_\sdG$ respectively are taken from some suitable probability distribution. 

Matrices $M_\sG$ follow the condition $M_\sG(j,k) = M_\sG(k,j)$, where $M_\sG(j,k) \in \mathbb{C}$. Thus a $N \times N$ $M_\sG$ matrix consist of $N(N+1)/2$ complex elements. 

Matrices $M_\G$ are the Ginibre matrix \cite{ginibre}, and do not have any symmetry. These consist of $N^2$ complex elements.

Matrices $M_\sdG$ are self-dual and consist of complex quaternion random numbers. For a quaternion number $Q = q_0 e_0 + q_1 e_1 + q_2 e_2 + q_3 e_3$ its dual $Q^D$ is defined as $Q^D = q_0 e_0 - q_1 e_1 - q_2 e_2 - q_3 e_3$, where the $q_j$ are the coefficient and the $e_j$ are its basis. In $M_\sdG$, the coefficients $q_j$ are complex numbers. A matrix is called self-dual if $M^D = M$, where $(M^D)_{jk} = \left(M_{kj}\right)^D$. It is more appropriate to write quaternion numbers in their $2 \times 2$ matrix representation, where the basis $e_0, e_1, e_2, e_3$ takes the form of the identity matrix, and the Pauli spin matrices, $e_0 = \mathbb{I}_2, e_1 = i \sigma_z, e_2 = -i \sigma_y, e_3 = i \sigma_x$. The quaternion number thus takes the form
\begin{align}
Q & =  q_0 e_0 + q_1 e_1 + q_2 e_2 + q_3 e_3 \\
& =  \left(
\begin{array}{rl}
	q_0 + i q_1 & q_2 + i q_3 \\
	-q_2 + i q_3 & q_0 - i q_1
\end{array}
\right ), \\
& = \left \{
\begin{array}{ll}
 \left(
\begin{array}{rl}
	z_1 & z_2 \\
	-z_2^\star & z_1^\star
\end{array}
\right ), & \text{for real quaternions,} \\
  \left(
\begin{array}{rl}
	z_1 & z_2 \\
	z_3 & z_4
\end{array}
\right ), & \text{for complex quaternions,}
\end{array} \right. 
\end{align}
where the $z_j \in \mathbb{C}$. In such a representation of quaternions, the matrix $M_\sdG$ takes the form of a $2N\times 2N$ matrix of complex numbers. Therefore, the matrix elements of $\mathcal M_\sdG$ (a $2N \times 2N$ representation of $M_\sdG$) satisfies following conditions,
\begin{equation}
\begin{aligned}
	\mathcal M_\sdG(2k-1, 2j-1) & = \mathcal M_\sdG(2j, 2k) \\
	\mathcal M_\sdG(2k-1, 2j) & = - \mathcal M_\sdG(2j-1, 2k) \\
	\mathcal M_\sdG(2k, 2j-1) & = -\mathcal M_\sdG(2j, 2k-1) \\
	\mathcal M_\sdG(2k, 2j) & = \mathcal M_\sdG(2j-1, 2k-1)
\end{aligned}
\end{equation}

We consider that the real and imaginary part of elements of $M_\sG, M_\G$, and all $8$ components of complex quaternions in $\mathcal M_\sdG$ follow the Gaussian distribution. The joint probability distribution (jpd) of these matrices can be written as,
\begin{align}
	\nonumber
	P(M)  = \exp\left(- \frac{\beta}{2}\text{Tr} M^\dag M \right),
\end{align}
where $\beta = 1, 2$ and $4$ for $M_\sG, M_\G$ and $M_\sdG$ respectively. For $\beta =4$, the jpd of $2N\times 2N$ representation, \textit{i.e.}, $\mathcal M_4$ follows the same distribution but with double the variance, \textit{i.e.},
	$P(\mathcal M_\sdG) = \exp(-(\beta/4) \text{Tr} \mathcal M^\dag_\sdG \mathcal M_\sdG)$

\begin{figure}
\centering
\includegraphics[width=0.85\linewidth]{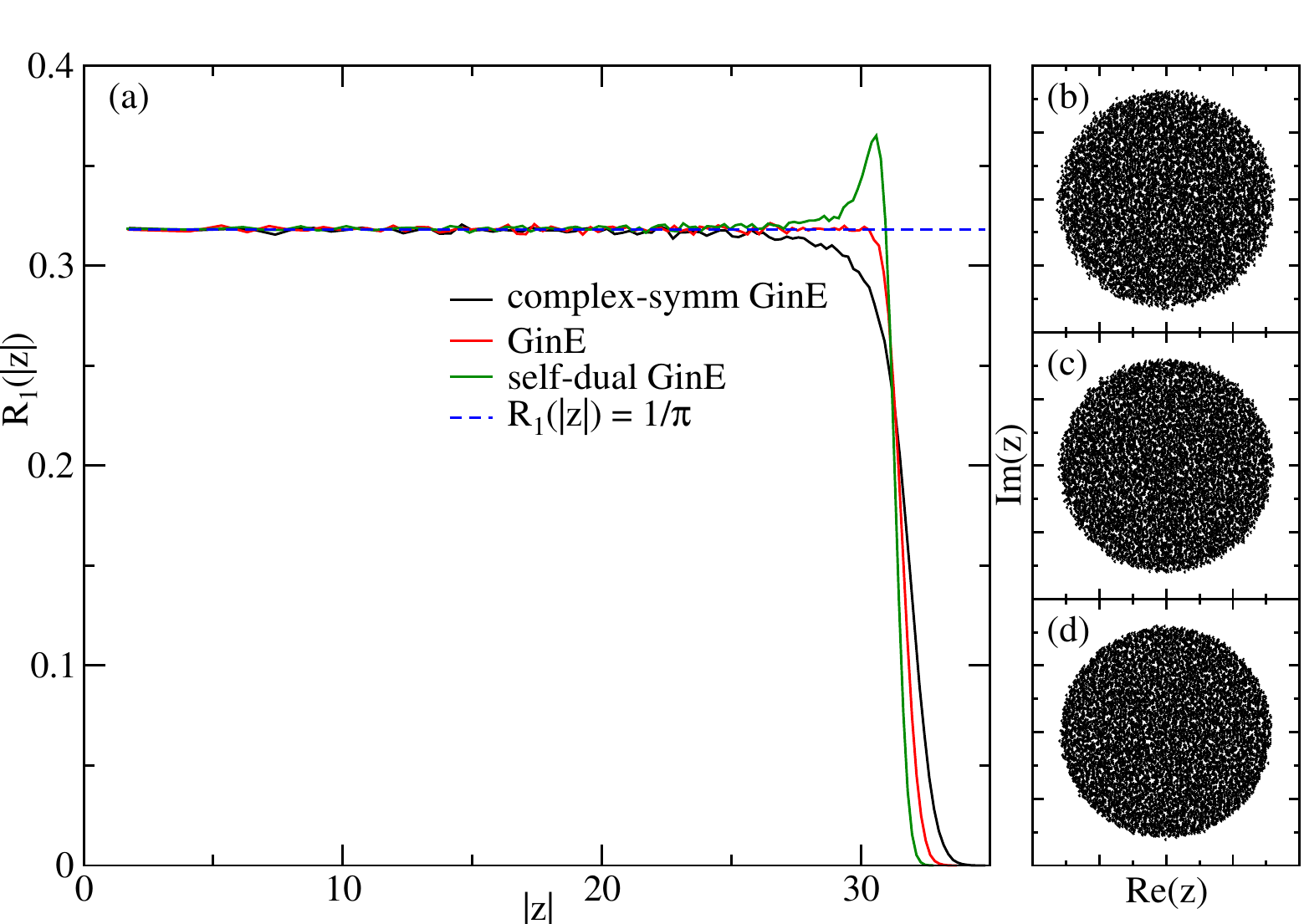}
\caption{The spectral density for all three ensembles are shown in (a). The scatter plots of eigenvalues for symm-GinE, GinE and self-dual-GinE are shown in (b), (c) and (d) respectively. \label{fig-r1-ev-all-rmt}}
\end{figure}

The matrices are diagonalized numerically to obtain their eigenvalues. For large $N$, the eigenvalues are distributed uniformly in a complex plane inside a disc of radius $\approx \sqrt{N}$ except for some variations at the boundary of the disc as shown in the Fig.~(\ref{fig-r1-ev-all-rmt}). Since matrix jpd is invariant under unitary transformations, the eigenvalue distribution is circularly symmetric. The spectral density or one-point correlation function $R_1(z) \equiv R_1(|z|)$, such that $\int R_1(|z|) d^2 z = N$, for all three ensembles, are also shown in the same figure. The spectral density is constant inside the disc. At the boundary, it decreases for $\beta = 1 ~(M_\sG)$, remains constant for $\beta = 2 ~(M_\G)$, and increases for $\beta = 4 ~(M_\sdG)$.  It suggests that eigenvalue repulsion is minimum for $\beta = 1$ and maximum for $\beta = 4$. Note that $\mathcal M_\sdG$ is self-dual and has degenerate eigenvalues. We ignore the degenerate eigenvalues in the scatter plot shown in Fig.~(\ref{fig-r1-ev-all-rmt}(d)).

\section{Eigenvalues Correlations}
Although matrix elements are distributed independently of each other, the eigenvalues are highly correlated. We are interested in the fluctuations of eigenvalues and their universality.
The spectral density can be categorized into two parts, an average and smooth distribution over a long range which depends on the distribution of matrix elements and remains the same across different matrices of the ensemble, and fluctuations of eigenvalues around the smooth and averaged distribution which is found to be universal in quantum chaotic systems
The fluctuation statistics of different systems can be compared only after re-scaling the spectra such that the global smooth variation in the spectral density of the rescaled spectrum becomes uniform and equal in each system. Such a transformation of eigenvalues is referred to as the unfolding of the spectrum.  We consider nearest neighbor spacing distribution, number variance, and spacing ratio to investigate universal aspects of fluctuations in unfolded spectra.

\section{Unfolding}
The unfolding refers to the transformation of eigenvalues so that the average spectral density of transformed eigenvalues is a constant. In one-dimensional spectra, the unfolding function is trivial to find as it transforms an elementary interval $dx$ having $R_1(x) dx$ average eigenvalues to $d\tilde x$ so that the average spectral density with respect to transformed eigenvalues, $\tilde x$, becomes one. Thus we have,
$d\tilde x = R_1(x) dx$
or $\tilde x_j = \int_0^{x_j} R_1(x) dx$.

For two-dimensional spectra obtained in symm-GinE, GinE, and self-dual-GinE, the spectral density is uniform and is equal to $R_1 = 1/\pi$. Therefore spacing distribution can be evaluated without any unfolding.

To compare fluctuation statistics of any other system having non-uniform density with that of the Ginibre ensemble, we unfold the eigenvalues to spectral density $R_1 = 1/\pi$. Similar to the unfolding of the one-dimensional spectrum, the elementary area in the two-dimensional spectrum is transformed as
\begin{align}
\label{eq-unfolding}
\frac{1}{\pi} d^2\tilde z = R_1(z) d^2 z,
\end{align}
where $\tilde z$ represents the unfolded spectra.

Above equation can have many solutions. However, It does not mean that unfolding is not unique. We emphasize that satisfying the condition of constant spectral density is not sufficient to unfold the spectrum. The unfolding is also supposed to preserve the ordering of the neighbors of an eigenvalue, \textit{i.e.}, the $k$-th nearest eigenvalue should remain the $k$-th nearest eigenvalue even after the unfolding.

We argue that the separation between two unfolded eigenvalues at a local scale is given by \cite{epl-2019, Ravi-2015}
\begin{align}
\label{eq-unfold-distance}
ds = \sqrt{\pi R_1(z)} |dz|,
\end{align}
where $|dz| = \sqrt{dr^2 + r^2 d\theta^2}$ in the polar coordinates. At a non-local scale, the separation between two eigenvalues in the unfolded spectrum is the distance along a curve (geodesics) so that $\int ds$ is minimum.

Eq.~(\ref{eq-unfold-distance}) represents a surface with the metric $ds^2 = \pi R_1(z) (dr^2 + r^2 d\theta^2)$. If the surface is curved, it is not possible to write the metric in a flat form like $ds^2 = d\tilde r^2 + \tilde r^2 d\tilde \theta^2$ and the unfolded eigenvalues lie on a curved surface. In some cases, where the metric $ds^2$ has zero curvature, the spectrum can be unfolded to a flat plane. We encounter one such case in the fluctuation statistics of random normal matrices where the spectral density is not uniform and the corresponding metric is flat. For completeness, it is discussed in detail in ref. \cite{supplementary}.

We unfold the spectrum locally to study spacing distribution. For number variance, the local unfolding can not work. We discuss in the section ahead how to unfold and evaluate the number variance when metric $ds^2$ has non-zero curvature.

\section{Local unfolding and spacing distribution}
\begin{figure}[!h]
\includegraphics[width=\linewidth]{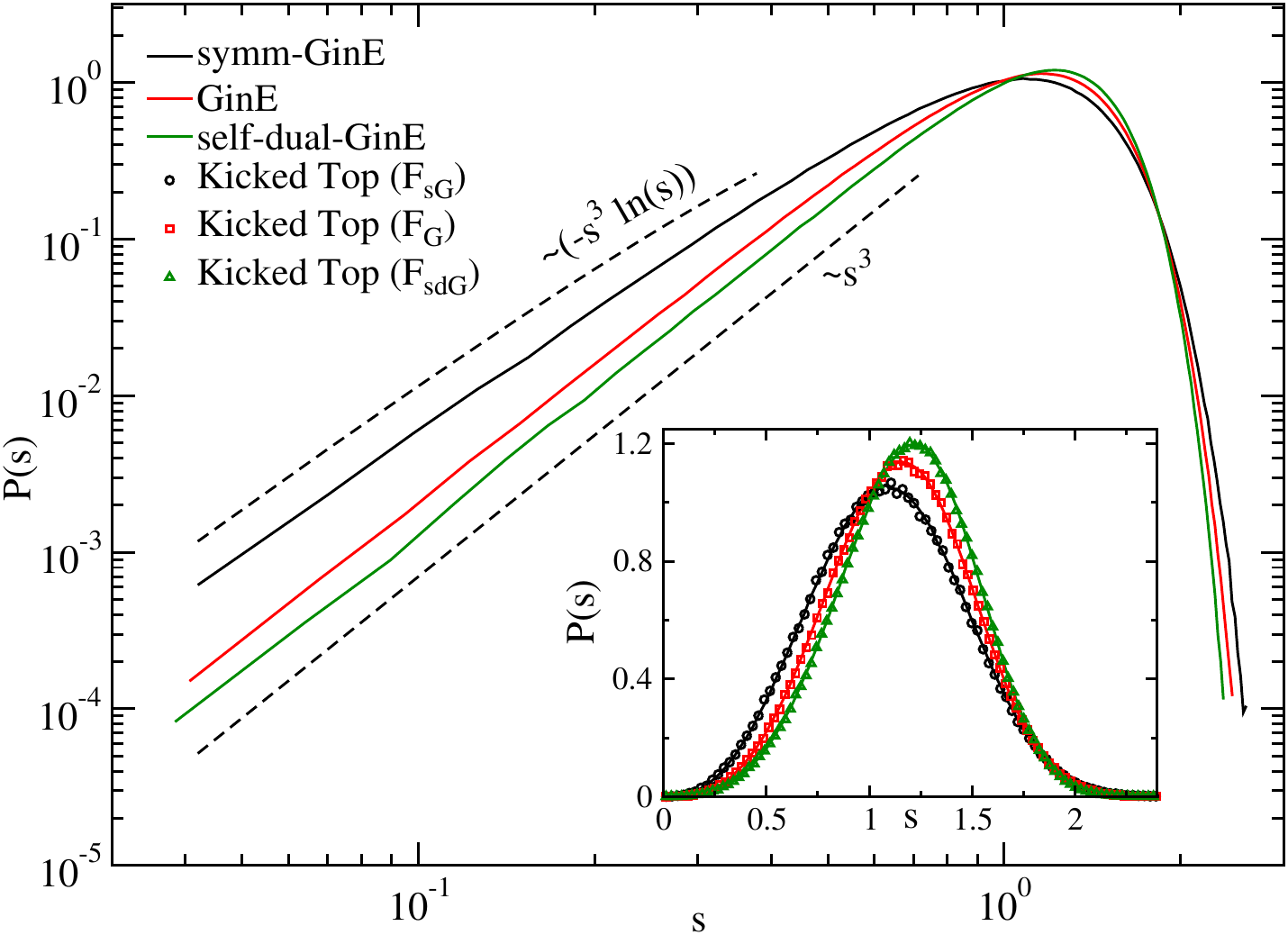}
\caption{Nearest neighbor spacing distribution for symm-GinE, GinE and self-dual-GinE on log-log scale and its agreement with the dissipative kicked top represented by $F_\sG, F_\G$ and $F_\sdG$ respectively. Same spacing distributions on linear scale are shown in the inset.\label{fig-sp-hist-rmt-dqktop-all}}
\end{figure}

The probability density $P(s)$ of having spacing $s$ between nearest neighbors for all three matrix ensembles are shown in Fig.~(\ref{fig-sp-hist-rmt-dqktop-all}), where we consider matrices with dimension $N = 1000$. We ignore the eigenvalues at the boundary where spectral density is not constant, therefore unfolding is not required. The spacing distribution shows a cubic repulsion $P(s) \propto s^3$ for small spacings for GinE $(M_\G)$ and self-dual-GinE $(M_\sdG)$ but shows a slightly weaker repulsion $P(s) \propto (-s^3 \log(s))$ for symm-GinE $(M_\sG)$. The spacing distributions for $N > 2$ dimensional matrices are different from that of two dimensional matrices \cite{epl-2019, Hamazaki-2020} but the order of repulsion for small spacings remains the same.

In order to confirm their universality, we consider a generic quantum system of spinning top with periodic and impulsive kick. The system can be easily modified to exhibit invariance under time reversal and rotation as required for orthogonal, unitary and symplectic ensembles. It is discussed ahead in more detail. The spectral density of the Floquet operator for this system is not uniform. The unfolded nearest neighbor spacing is evaluated by multiplying each pair of nearest neigbhor by $\sqrt{\pi R_1(z)}$, where $R_1(z)$ represents the average density around the pair of eigenvalues. The local unfolding is sufficient to study nnsd as spectral density does not vary significantly. However, it may not work for $k$-th nnsd with large $k$. The spacing distributions show an excellent agreement with corresponding random matrix ensembles, as shown in Fig.~(\ref{fig-sp-hist-rmt-dqktop-all}).

\section{Spacing ratio}
In quantum chaotic systems with one-dimensional spectra, the ratio $r_k = (x_{k+1} - x_{k})/(x_{k+2} - x_{k+1})$ where the $x_k$ are eigenvalues in ascending order, follows a universal distribution. The spacing ratio is preferred over spacing distribution because it does not need the unfolding of the spectra (if the average spectral density is approximately the same in the region $x_k$ to $x_{k+2}$, the rescaling term being the same in both numerator and denominator does not alter the spacing ratio for folded and unfolded spectrum).

\begin{figure}[!h]
\centering
\includegraphics[width=0.8\linewidth]{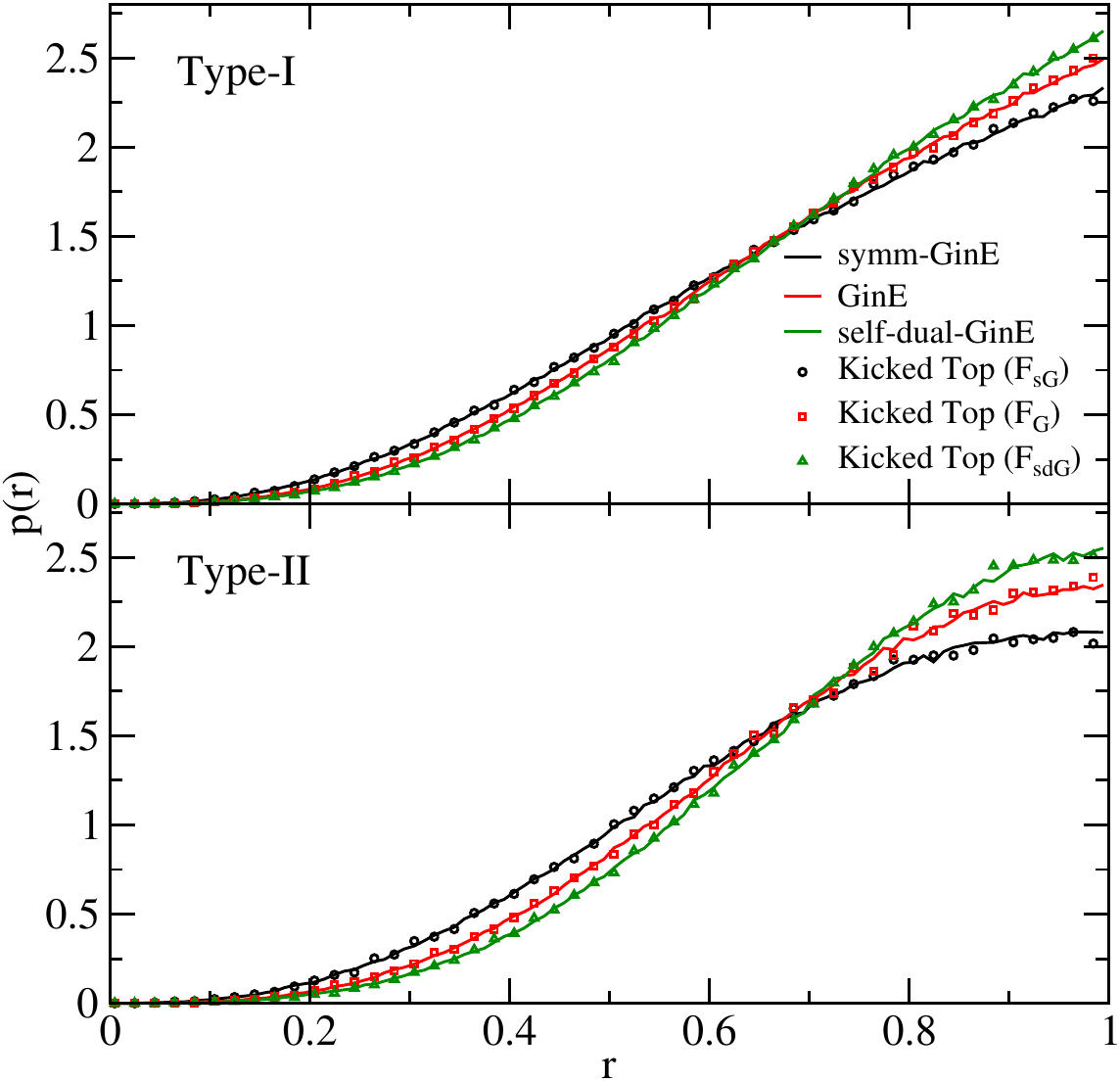}
\caption{Distribution of the spacing ratio of both types for random matrix ensembles represented by $M_\sG, M_\G$ and $M_\sdG$. The spacing ratios for dissipative kicked tops in Eq.~(\ref{eq-F-dissp}) show an excellent agreement with that of corresponding random matrix ensembles. \label{fig-sp-ratio}}
\end{figure}

For eigenvalues in the complex plane, we define the spacing ratio of two types. In type-I, we calculate the ratio $r = AB/AC$ where $B$ and $C$ are the nearest and next nearest neighbor of an eigenvalue $A$ \cite{Dusa-2022, Lucas-2020, epl-2019}. In type-II, we calculate the ratio $\min(r, 1/r)$ with $r = AB/BC$ where $B$ and $C$ represent the nearest neighbor of $A$ and $B$ respectively, excluding the cases where two eigenvalues are the nearest neighbor of each other \cite{epl-2019}. The distribution of the spacing ratio for all three ensembles are shown in Fig.~(\ref{fig-sp-ratio})

We also evaluate the spacing ratio in the same figure for the open quantum kicked top to verify the universality of the spacing ratio. We find an excellent agreement with the spacing ratio of quantum kicked top, which again validates the universality of spectral correlations in random matrix ensembles represented by, $M_\sG, M_\G$ and $M_\sdG$.

\section{Non-local unfolding and number variance}
The spacing distribution is useful for understanding the correlations of the shortest length. Number variance is more appropriate to study the correlation of larger lengths. The number variance measures the fluctuation around mean values of numbers of eigenvalues in a given region. If a spectrum in the complex plane with uniform density is divided into several regions of identical shape and size, the numbers of eigenvalues in each disc fluctuate around its mean values. If $n_1, n_2,...n_k$ are the numbers of eigenvalues in each disc, the number variance is defined as
\begin{align}
\label{eq-nvar-def}
\Sigma^2(\langle n \rangle) = \langle n^2 \rangle - \langle n\rangle^2,
\end{align}
where $\langle n^2 \rangle  = (\sum_{j=1}^k n_j^2)/k$ and $\langle n\rangle = (\sum_{j=1}^k n_j)/k$.
The number variance also exhibits universality and is affected by the symmetry of the system only.

For the spectrum in the complex plane, we count eigenvalues inside circular discs to measure the number variance. The radius is $r = \sqrt{\langle n \rangle}$ so that it accommodates $\int R_1(z) d^2z = \langle n \rangle$ eigenvalues. For uncorrelated systems, \textit{e.g.}, random points in the complex plane, since the two-point correlation $R_2(z_1,z_2) = R_1(z_1) R_1(z_2)$, it is easy to prove that the number variance grow linearly, $\Sigma^2(\langle n \rangle) = \langle n \rangle$ for small $\langle n \rangle$ \cite{Garcia-2023}. For the Ginibre ensemble, since eigenvalue jpd and all the correlations are known \cite{ginibre, mlmehta}, we find analytic expression of number variance for GinE as given in Eq.~(\ref{nmcr-sigmasq-disc}). The details of the derivation are given in the appendix.
\begin{align}
\label{nmcr-sigmasq-disc}
\Sigma^2_\G\left(\langle n \rangle\right) = \langle n \rangle -\int_0^{\langle n \rangle} \int_0^{\langle n \rangle}  I_0(2\sqrt{\eta_1 \eta_2}) e^{-\left(\eta_1 + \eta_2\right)} ~ \mathrm d\eta_1 \mathrm d\eta_2,
\end{align}
where $I_0(x) = \sum_{k=0}^\infty x^{2k}/(2^{2k}k!\Gamma(k+1))$ is the modified Bessel function of the first kind.

\begin{figure}[!h]
\centering
\includegraphics[width=0.9\linewidth]{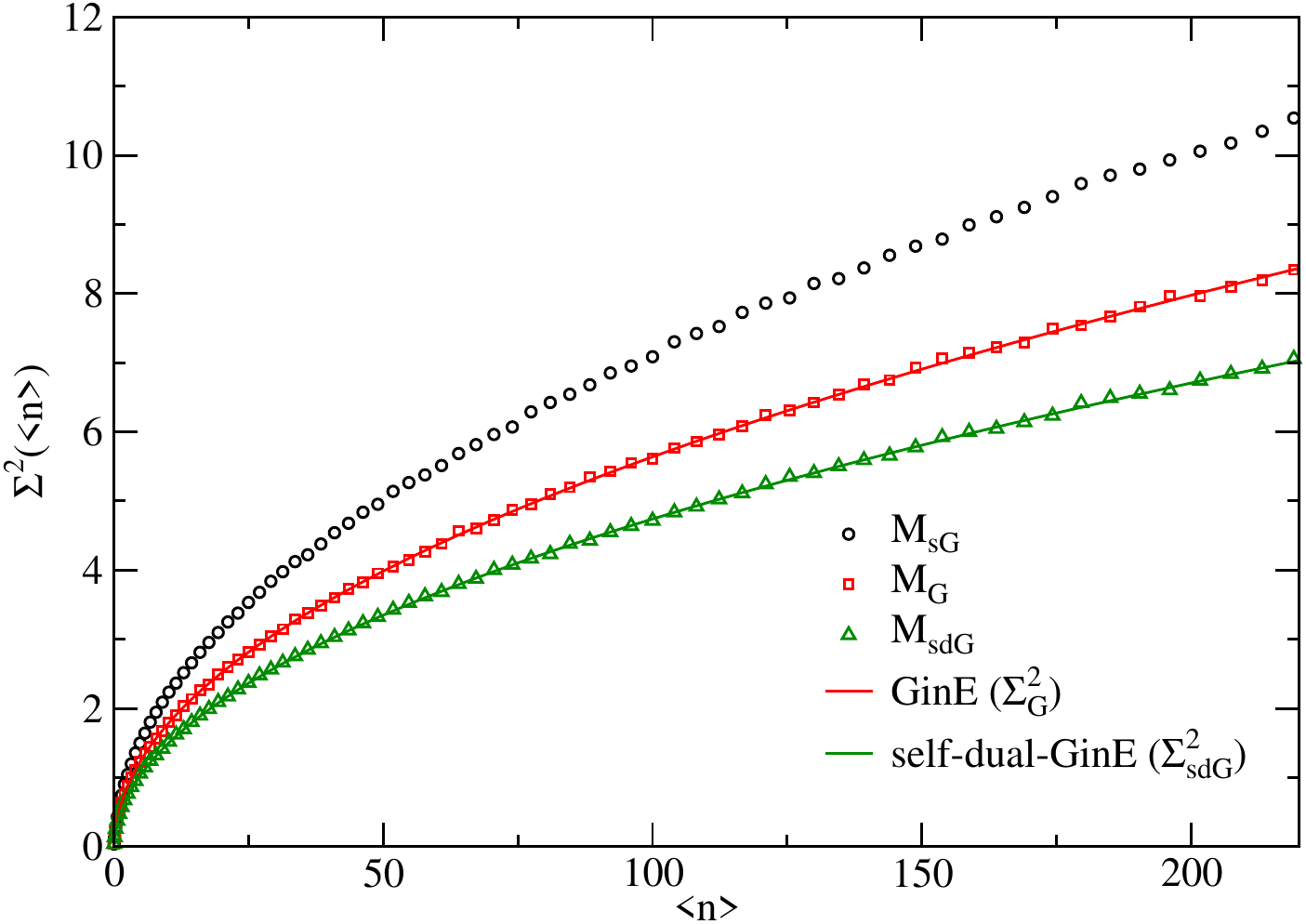}
\caption{Number variance for three ensembles viz., symm-GinE ($M_\sG$), GinE ($M_\G$) and self-dual-GinE ($M_\sdG$). Number variance for GinE follow Eq.~(\ref{nmcr-sigmasq-disc}) and that of self-dual-GinE shows excellent agreement with Eq.~(\ref{eq-nvar-self-dual-GinE}). \label{fig-var}}
\end{figure}

The eigenvalue jpd for other two ensembles, symm-GinE and self-dual-GinE, are not known to the best of our knowledge \cite{Hastings-2001}. We numerically find a simple expression for the number variance for self-dual-GinE, given by
\begin{equation}
\label{eq-nvar-self-dual-GinE}
\Sigma^2_\sdG \left(\langle n \rangle\right) = \Sigma^2_\G \left(\langle n \rangle/ \sqrt{2}\right)
\end{equation}
which shows an excellent agreement with numerical results for $\langle n \rangle > 1$.
The number variance for all three ensembles is shown in Fig.~(\ref{fig-var}).
Unlike spacing distributions, number variance show significant differences for these three ensembles, which makes it more suitable to study correlations in dissipative systems.

For the spectra with non-uniform spectral density, the number variance can be evaluated only after unfolding the spectra. The spectra of dissipative kicked tops with Floquet operators given by Eq.~(\ref{eq-F-dissp}) are circularly symmetric and spectral density depends only on $|z|$. The spectral density is obtained through a polynomial fit on numerical data. For such cases, it is very unlikely that the metric $ds^2$ has zero curvature (or unfolded spectra lie in the complex plane).

The number variance is calculated numerically as follows. Instead of finding unfolded eigenvalues, we look for the closed curve in actual (folded) spectra equivalent to circular disc. We consider an arbitrary point in the actual (folded) spectra as center and find the closed curve that have distance $s = \sqrt{\left< n\right>}$ from the point along geodesics in the complex plane, where the infinitesimal length is defined as in Eq.~(\ref{eq-unfold-distance}). We first find geodesics starting from the chosen point in all directions. Using the variational principle, the differential equation followed by geodesics in the polar coordinates can be written as,
\begin{align}
\label{eq-geodesics}
\frac{d}{d\theta} \frac{\partial}{\partial \dot r} L(r, \dot r, \theta) - \frac{\partial}{\partial r} L(r, \dot r, \theta) = 0,
\end{align}
where $L = \sqrt{\pi R_1(|z|)} \sqrt{r^2 + {\dot r}^2}$ with $\dot r = dr/d\theta$. The locus of the end point of these geodesics constitute the desired closed curve. We calculate numbers of eigenvalues in these closed curves and substitute them in Eq.~(\ref{eq-nvar-def}) to evaluate the number variance. The number variance for kicked top represented by the Floquet operators in Eq.(\ref{eq-F-dissp}) is shown in Fig.~(\ref{fig-var-ktop}). It shows good agreement with random matrix ensembles consistent with the symmetries of the top.

\section{Quantum Kicked Top}
Consider the Hamiltonian consisting of components of angular momentum,
\begin{equation}
H = \alpha J_x + \tau J_z^2 \sum_{n = -\infty}^\infty \delta(t-n).
\end{equation}
The first term represents a free precession around $J_x$ with velocity $\alpha$ followed by an impulsive force which produces a non-uniform rotation around $J_z$ in every period. Since $[H, J] = 0$, total angular momentum is the only conserved quantity. The top exhibits chaotic behavior as $\tau$ increases. The stroboscopic map over a period is appropriate to study the dynamics. The Floquet operator defined as the unitary evolution over a period $\mathcal F = \exp(\int_\mathcal T i H dt)$ has its eigenvalues on a unit circle. The fluctuation statistics of eigen-angles are similar to that of the circular ensemble of random matrices for chaotic systems. For the above system the Floquet operator,
\begin{equation}
	\label{eq-F-OE}
\mathcal F_\text{OE} =  e^{-i \tau J_z^2} e^{-i\alpha J_x},
\end{equation}  
belongs to the orthogonal class because it is invariant under the time reversal operator with $T =\exp(i \alpha J_x) K$, where $K$ is the complex conjugation operator. Note that it is not conventional time reversal in the sense that $T J T \ne -J$ but it reverses the dynamics, \textit{i.e.}, $T \mathcal F T^{-1} = \mathcal F^\dagger$. Since $T^2 = +1$, the spectral correlations of eigen-angles of $F$ are the same as that of the COE.

The kicked top having correlation similar to that of the CUE is obtained by breaking the time-reversal symmetry. The Floquet operator is given by
\begin{align}
\label{eq-F-UE}
\mathcal F_\text{UE} = e^{-i k J_y^2/(2J)} ~e^{-i \tau J_z^2/(2J)} ~e^{- i\alpha J_x}.
\end{align}

Similarly following Floquet operator is invariant under time reversal with $T^2 = -1$ for half-integer values of $J$. It does not have any rotational invariance and therefore exhibits the same spectral correlation as that of the CSE.
\begin{align}
\label{eq-F-SE}
\mathcal F_\text{SE} = e^{-\frac{i \tau_1 J_z^2}{2J}}  e^{-\frac{i}{2J}\left(\tau_2 J_z^2 + \tau_3 (J_xJ_z+J_zJ_x) + \tau_4(J_x J_y + J_yJ_x)\right)},
\end{align}
The time reversal operator for $\mathcal F_{\text{SE}}$ is given by $T = \exp\left(-i \tau_1 \frac{J_z^2}{2J}\right) K$.

\section{The dissipative kicked top}
The dissipation causes the momentum to decrease with time. To describe a dissipative kicked top, we operate a dissipation operator defined as $D = \exp(-\gamma J_z^2/2J)$, where $\gamma$ is a control parameter. We show that the modified form of Floquet operators, $\mathcal F_\text{OE}, \mathcal F_\text{UE}$ and $\mathcal F_\text{SE}$ exhibit the same spectral statistics as that of symm-GinE, GinE, and self-dual-GinE respectively. The Floquet operator for dissipative tops, $F$, can be written as
\begin{align}
\label{eq-F-dissp}
F_\sG = D \mathcal F_\text{OE},~~~F_\G = D \mathcal F_\text{UE},~~~F_\sdG = D \mathcal F_\text{SE}
\end{align}

The Floquet operators are no longer unitary. Their eigenvalues start falling towards the center and form a ring-like structure. To numerically simulate the eigenvalues, spectral density, and other relevant statistics, we consider ensembles of spectrum by varying one or two parameters. The order of variation is kept much smaller than the parameter values to reduce the variation in the spectral density.

\begin{figure}[!h]
\centering
\includegraphics[width=\linewidth]{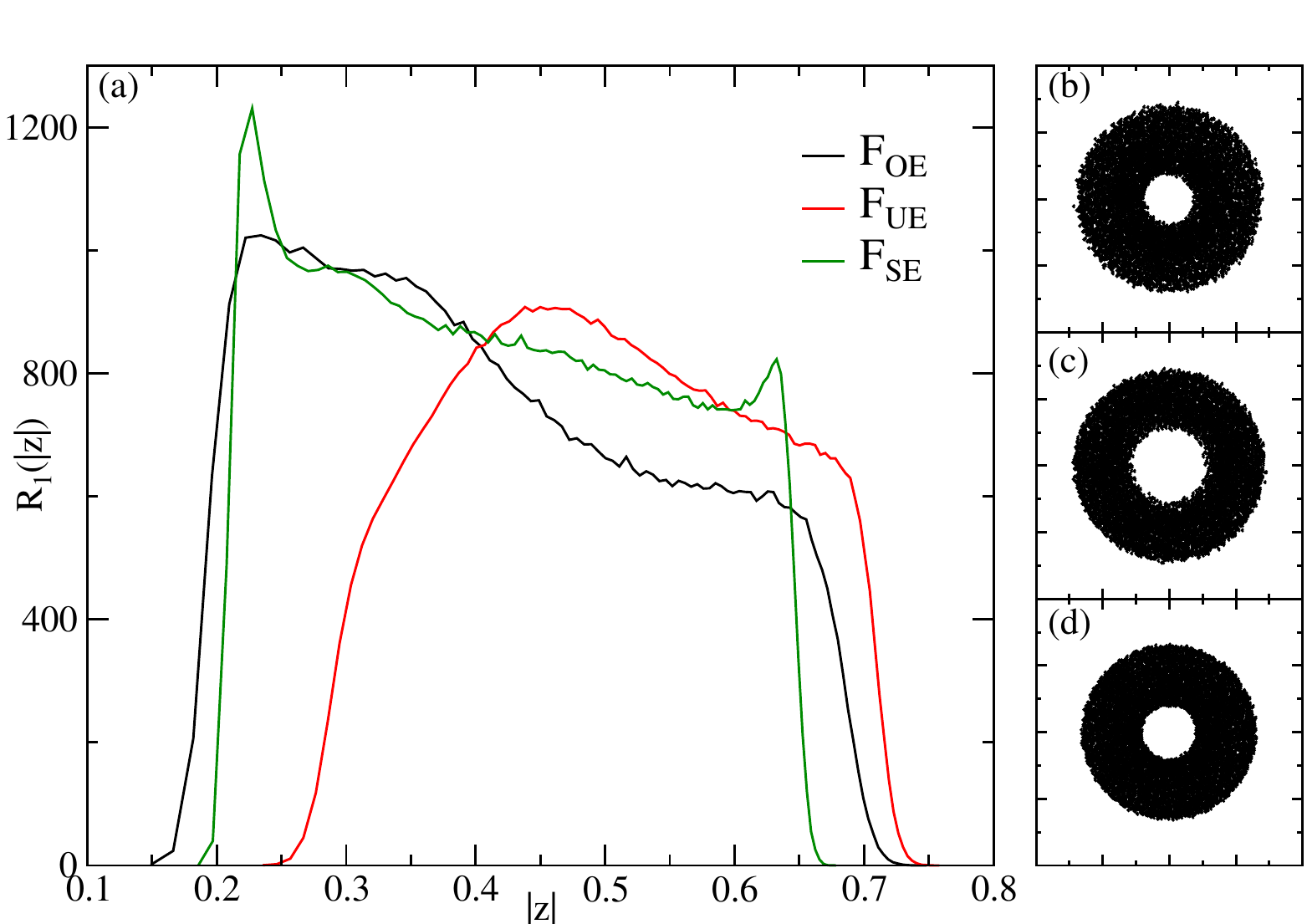}
\caption{(a) The spectral density for the eigenvalues of the Floquet operators given by Eq.~(\ref{eq-F-dissp}). The scatter plot of eigenvalues for $F_\sG, F_\G$ and $F_\sdG$ are shown in (b), (c) and (d) respectively. \label{fig-r1-ev-all-dqktop}}
\end{figure}

We consider the following parameters to numerically calculate eigenvalues. For eigenvalues of $F_\sG$, we substitute $\gamma = 5/N, \alpha = 7$ and $300 \le \tau \le 350$ in Eqs.~(\ref{eq-F-OE}, \ref{eq-F-dissp}). The eigenvalues of $F_\G$ are calculated by substituting $\gamma = 4/N, \alpha = 25, 40 \le \tau \le 50$ and $60 \le k \le 66$ in Eqs.~(\ref{eq-F-UE}, \ref{eq-F-dissp}). Finally for $F_\sdG$, we substitute $\gamma = 5/N, \tau_1 = 307, \tau_2 = 336, 506 \le \tau_3 \le 530$ and $370 \le \tau_4 \le 420$ in Eqs.~(\ref{eq-F-SE}, \ref{eq-F-dissp}). Here $N = 2J + 1$ with $J = 1000$ in expression for $F_\sG$ and $F_\G$, and $J = 999.5$ in $F_\sdG$. The scatter plot of eigenvalue and average spectral density $R_1(r)$, such that $\int 2\pi r R_1(r) dr = N$, are shown in Fig.~(\ref{fig-r1-ev-all-dqktop}). Note that the spectrum of $F_\sdG$ is doubly degenerate due to the Kramer's degeneracy \cite{Scharf-1988}. We ignore the degenerate eigenvalues in all calculations.

\begin{figure}[!h]
	\centering
	\includegraphics[width=0.9\linewidth]{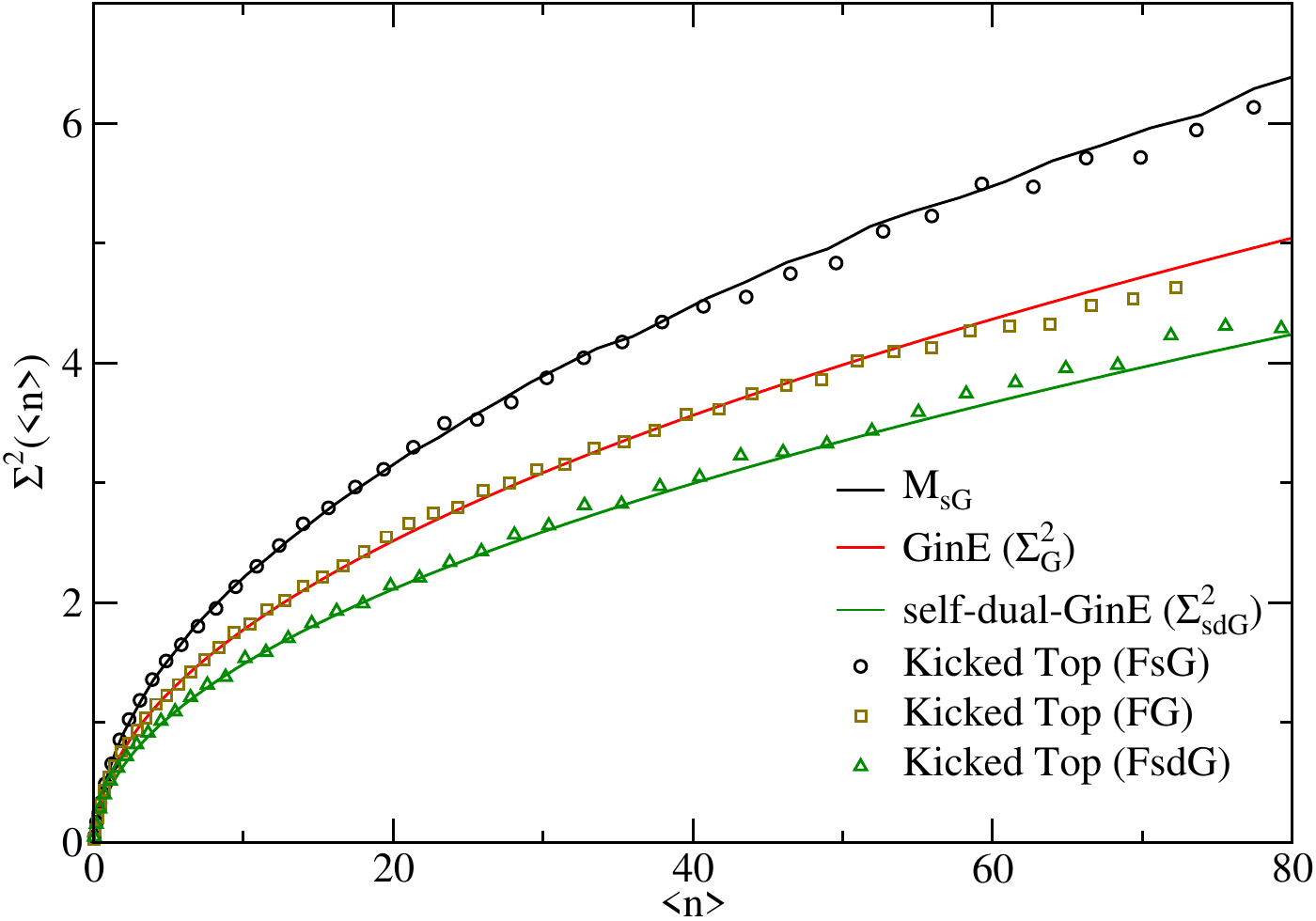}
	\caption{Number variance for the dissipative top defined by Floquet operators in Eq.~(\ref{eq-F-dissp}) and their agreement with corresponding random matrix ensembles.\label{fig-var-ktop}}
\end{figure}

The nearest neighbor spacing distribution after local-unfolding, spacing ratio, and number variance after non-local unfolding are shown in Fig.~(\ref{fig-sp-hist-rmt-dqktop-all},\ref{fig-sp-ratio})  and (\ref{fig-var-ktop}) respectively. We ignore the eigenvalues near the inner and outer boundaries to avoid any finite-dimensional effect. An excellent agreement of all fluctuation statistics validates that the spectral correlations in symm-GinE, GinE, and self-dual-GinE are universal and found in dissipative chaotic systems with TRI and RI, without TRI and RI, and with TRI but without RI respectively. 

\section{Conclusion}
We discuss the short-range and long-range correlations in terms of spacing distribution, spacing ratio, and number variance statistics for symm-GinE, GinE, and self-dual-GinE. We have shown that quantum chaotic systems with symmetry class as that of orthogonal (GOE or COE), unitary (GUE or CUE), and symplectic (GSE or CSE) ensembles in the presence of dissipative environment show the same spectral correlations as that of above-mentioned ensembles.

We have also studied the method to perform unfolding at non-local scale for spectra in the complex plane with non-uniform density and verified the universality of long-range correlations in terms of number variance for a generic system of dissipative kicked top. Unlike spacing distribution which shows the same cubic repulsion for all three ensemble, the number variance show significant variations. We also find a simple relation for number variance between GinE and self-dual-GinE. We believe that this letter will be useful to understand long-range correlations in terms of number variance, spectral rigidity, and form factor for systems with two-dimensional spectrum in various integrable and non-integrable problems.

\renewcommand{\theequation}{A\arabic{equation}}
\setcounter{equation}{0}
\section{Appendix}

The averages $\langle n \rangle$ and $\langle n^2\rangle$ in Eq.~(\ref{eq-nvar-def}) can be written as,
$\left<n\right>  = \int R_1(z) ~\mathrm d^2 z$ and 
$\left<n^2\right>  = \int \int  R_2(z_1,z_2)~ \mathrm d^2z_1 \mathrm d^2z_2 + \int R_1(z) ~\mathrm d^2 z$. 
Here last term in the second equation represents the self-correlation. The spectral density and two-eigenvalue correlation for the Ginibre ensemble are given by $R_1(z) = 1/\pi$ and $R_2(z_1,z_2) = \frac{1}{\pi^2} \left(1 - \exp\left(-|z_1 -z_2|^2\right)\right)$ respectively \cite{ginibre, mlmehta, haakebook}. Substituting above expressions in Eq.~(\ref{eq-nvar-def}) and integrating in a circular region of radius $r = \sqrt{\langle n \rangle}$ we get
\begin{eqnarray}
\label{nmcr-sigmasq}
 \Sigma^2(\langle n \rangle)  =\int \frac{1}{\pi} ~\mathrm d^2z
- \frac{1}{\pi^2}\iint e^{- |z_1 - z_2|^2} ~\mathrm d^2z_1 \mathrm d^2z_2.
\end{eqnarray}
The Eq.~(\ref{nmcr-sigmasq-disc}) of the main text is obtained after simplifying above equation.


\acknowledgments


R.P. thanks Dushyant Kumar and Richa Arya for useful discussion.

\bibliography{refs}

\end{document}


\begin{center}
		\textbf{Supplementary material for ``Universality of spectral fluctuations in open quantum chaotic systems"}\\
	Jisha C, Ravi Prakash
\end{center}

In this supplementary material we evaluate the number variance for the eigenvalue jpd given by
\begin{align}
\label{eq-jpd}
P(z_1, z_2, \ldots, z_N) \propto \prod_{1 \le j < k \le N} |z_j - z_k|^2 e^{-\sum_{l = 1}^N V(|z_l|)}.
\end{align}
Such correlations are found in the eigenvalues of random matrices which commute with their hermitian conjugate, also known as normal matrices \cite{Ravi-2015}. For the potential $V(|z|) = |z|^2$, the jpd corresponds to the Ginibre ensemble. We have proved analytically in ref.~\cite{Ravi-2015} that all $n$-eigenvalue correlations after unfolding exhibit universality and become independent of the choice of the potential $V$.

We shall find a map that unfolds the spectra at a non-local scale for $V(|z|) = |z|^{2k}$ and show that the number variance of the unfolded spectra is the same as that of the Ginibre ensemble. The spectral density for the jpd in Eq.~(\ref{eq-jpd}) for $V(|z|) = |z|^{2k}$ can be written as \cite{Ravi-2015},
\begin{equation}
\label{eq-r1}
R_1(|z|) = \frac{1}{\pi}k^2 |z|^{2(k-1)}.
\end{equation}

\begin{figure}[!h]
\includegraphics[width=0.5\linewidth]{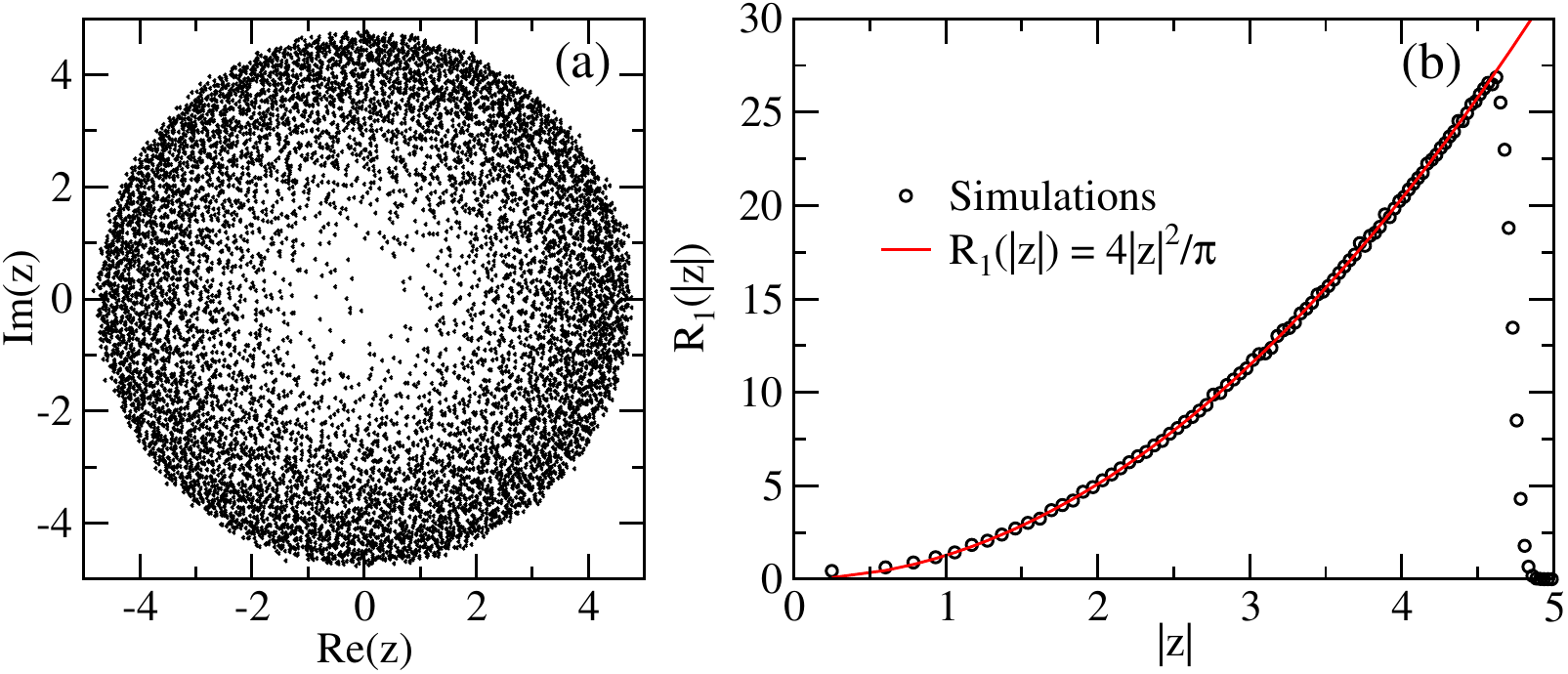}
\caption{(a) The eigenvalue scatter plot, and (b) the spectral density for the jpd in Eq.~(\ref{eq-jpd}) for $V(|z|) = |z|^4$. \label{fig-ev-r1}}
\end{figure}

We use the Monte-Carlo method to numerically generate eigenvalues for the jpd in Eq.~(\ref{eq-jpd}) for $V(|z|) = |z|^{2k}$ with $k = 2$ \cite{Ravi-2015}. The eigenvalues scatter plot and the spectral density are shown in Fig.~(\ref{fig-ev-r1}).

It can be easily shown that the metric $ds^2 = \pi R_1(|z|) |dz|^2$, where $|dz|^2 = dr^2 + r^2 d\theta^2$ in the polar coordinates is flat for the spectral density given in Eq.~(\ref{eq-r1}), and it can be written in a flat form as $ds^2 = |d\tilde{z}|^2 = d\tilde r^2 + \tilde r^2 d\tilde \theta^2$, where $\tilde z = \tilde r \exp(i \tilde \theta)$. Thus unfolding functions, $\tilde r  = \tilde r (r,\theta)$ and $\tilde \theta = \tilde \theta(r,\theta)$ can be derived from
\begin{align}
\label{eq-unfold}
d\tilde r^2 + \tilde r^2 d\tilde \theta^2 =  \pi R_1(r)(dr^2 + r^2 d\theta^2).
\end{align}

Eq.~(\ref{eq-unfold}) implies that
\begin{align}
\left(\frac{\partial \tilde r}{\partial r} \right)^2 + \tilde r^2 \left(\frac{\partial \tilde \theta}{\partial r} \right)^2 & = \pi R_1(r) \\
\left(\frac{\partial \tilde r}{\partial \theta} \right)^2 + \tilde r^2 \left(\frac{\partial \tilde \theta}{\partial \theta} \right)^2 & = \pi r^2 R_1(r) \\
\frac{\partial \tilde r}{\partial r} \frac{\partial \tilde r}{\partial \theta} + \tilde r^2 \frac{\partial \tilde \theta}{\partial r}  \frac{\partial \tilde \theta}{\partial \theta} & = 0
\end{align}
We look for a solution separable in $r$ and $\theta$, \textit{i.e.}, $\tilde r = \tilde r (r)$ and $\tilde \theta = \tilde \theta (\theta)$. We obtain one such solution given by
\begin{align}
\label{eq-unfolding-function}
\tilde r = r^k~\text{and}~\tilde \theta  = k \theta.
\end{align}

\begin{figure}[!h]
	\includegraphics[width=0.5\linewidth]{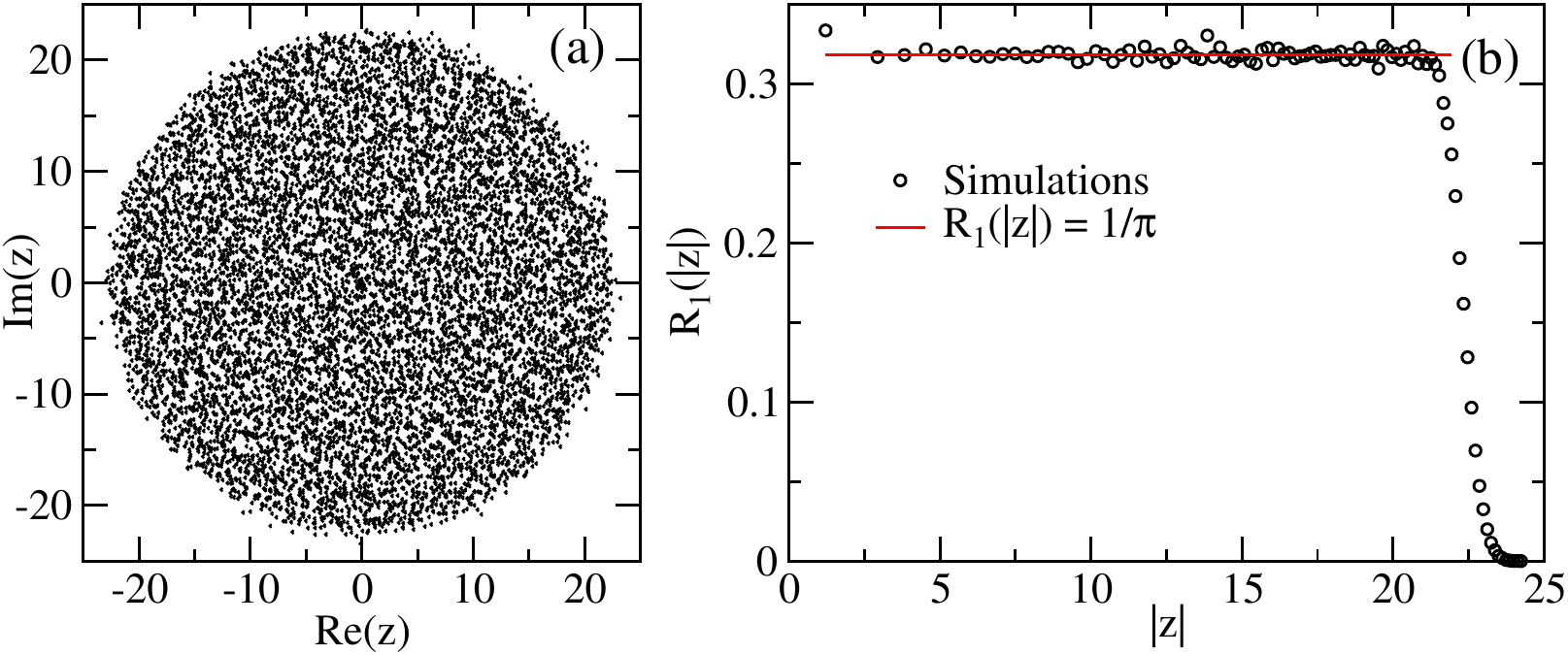}
	\caption{(a) Eigenvalue scatter plot, and (b) the spectral density for the spectra unfolded using Eq.~(\ref{eq-unfolding-function}) for $V(|z|) = |z|^4$ (\textit{i.e.}, $k = 2$). \label{fig-ev-r1-unfold}}
\end{figure}

We unfold the spectra shown in Fig.~(\ref{fig-ev-r1}) using Eq.~(\ref{eq-unfolding-function}). The scatter plot and the spectral density of unfolded eigenvalues are shown in Fig.~(\ref{fig-ev-r1-unfold}). Note that the unfolded spectra lies on two Riemann sheets for $k = 2$.

\begin{figure}[!h]
	\includegraphics[width=0.5\linewidth]{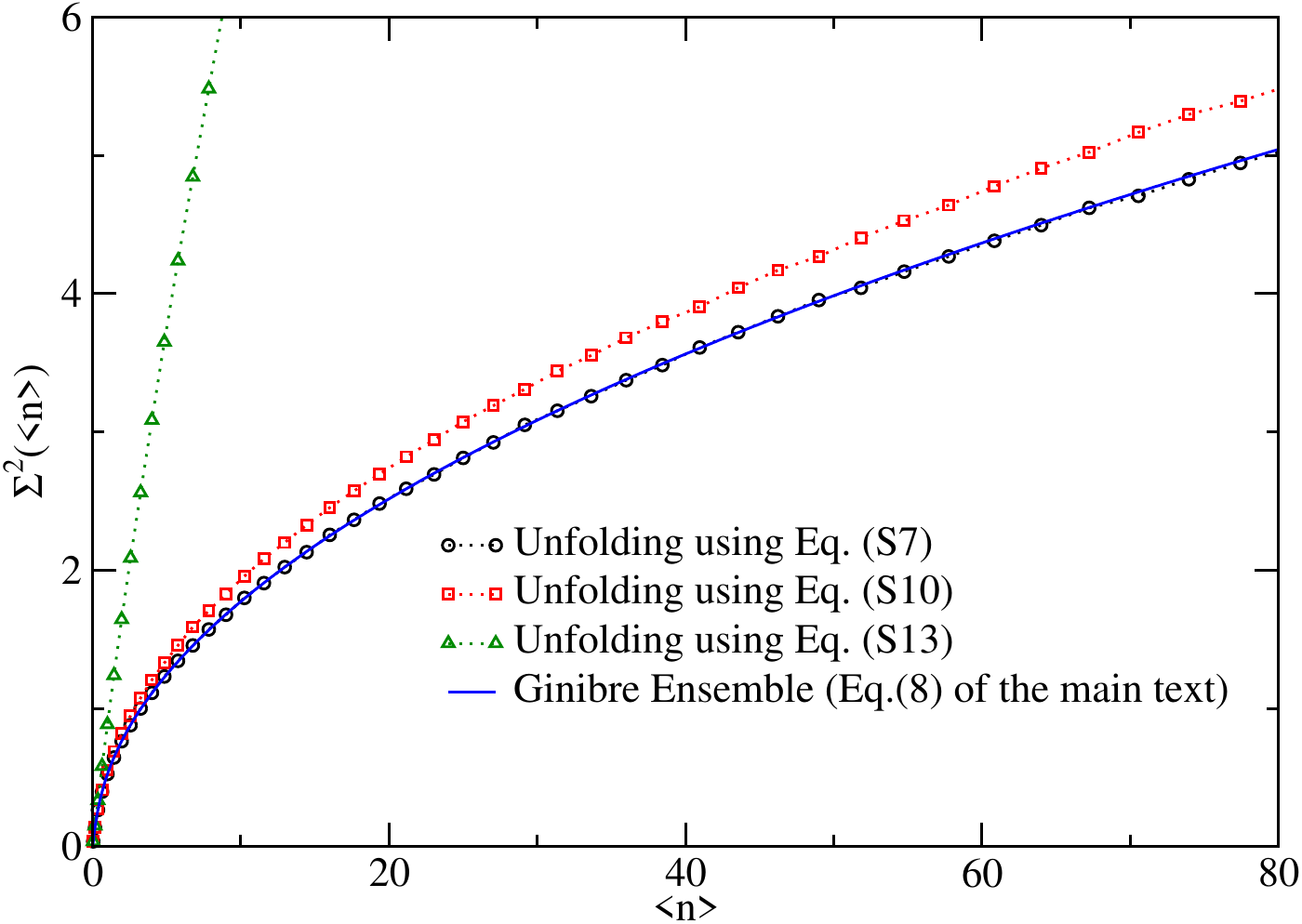}
	\caption{The spectra obtained from the jpd in Eq.~(\ref{eq-jpd}) with $V(|z|) = |z|^4$ is unfolded using three methods. We plot number variance of the same spectra for three types of unfolding functions given by Eqs.~(\ref{eq-unfolding-function}), (\ref{eq-unfolding-function-2a}) and (\ref{eq-unfolding-function-3a}). \label{fig-nvar-all}}
\end{figure}

The unfolded spectra has density $R_1 = 1/\pi$. To calculate number variance, we count eigenvalues inside circles of radius $r = \sqrt{\langle n \rangle}$ placed at arbitrary positions in one of the Riemann sheets, and measure their variance. The number variance for the unfolded spectra is shown in Fig.~(\ref{fig-nvar-all}). It shows an excellent agreement with the number variance for the Ginibre ensemble, given by Eq.~(8) of the main text.

\subsection{Some unfoldings exhibiting incorrect fluctuation statistics}
\begin{figure}[!h]
	\includegraphics[width=0.5\linewidth]{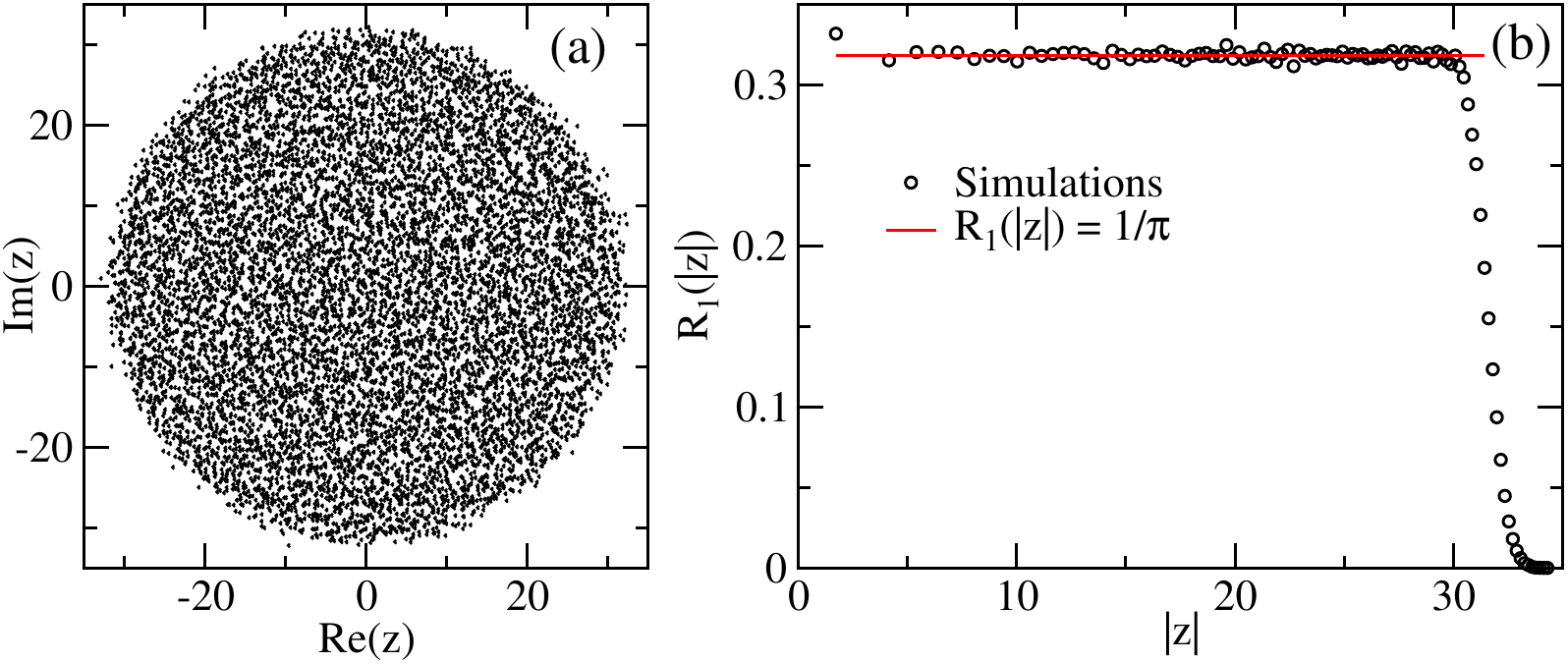}
	\caption{(a) The eigenvalue scatter plot, and (b) the spectral density for the spectra in Fig.~(\ref{fig-ev-r1}) unfolded using Eq.~(\ref{eq-unfolding-function-2a}). \label{fig-ev-r1-unfold-2}}
\end{figure}

For radially symmetric spectral density $R_1(z) = R_1(|z|)$, a common approach to unfold the spectrum is to consider it effectively one dimensional and unfold only the radial component of eigenvalues \cite{Garcia-2023}. Eq.~(5) of the main text can also be written in the polar coordinates as
\begin{align}
	\label{eq-unfolding-function-2}
	d\tilde \theta & = d\theta, \\
	\text{and}~ \frac{1}{\pi} \tilde r d \tilde r & = R_1(r) r dr.
\end{align}
We get following unfolding function,
\begin{align}
		\label{eq-unfolding-function-2a}
\tilde \theta = \theta 
~\text{and}~\tilde r = r^2 \sqrt{2}.
\end{align}
We show that such unfolding may lead to erroneous results. We unfold the spectra for the jpd in Eq.~(\ref{eq-jpd}) for $V(|z|) = |z|^4$ using the above transformation. The eigenvalue scatter plot and the spectral density are shown in Fig.~(\ref{fig-ev-r1-unfold-2}).
See Fig.~(\ref{fig-nvar-all}) for the number variance of the unfolded spectra. It shows significant differences from that of the Ginibre ensemble. Such an unfolding, in general, does not guarantee that fluctuations are preserved. The transformation moves the eigenvalues along the radial direction to achieve a constant spectral density but in this process, eigenvalues have changed the ordering of their neighbors.

\begin{figure}[!h]
	\includegraphics[width=0.5\linewidth, trim = 1 210 10 60, clip]{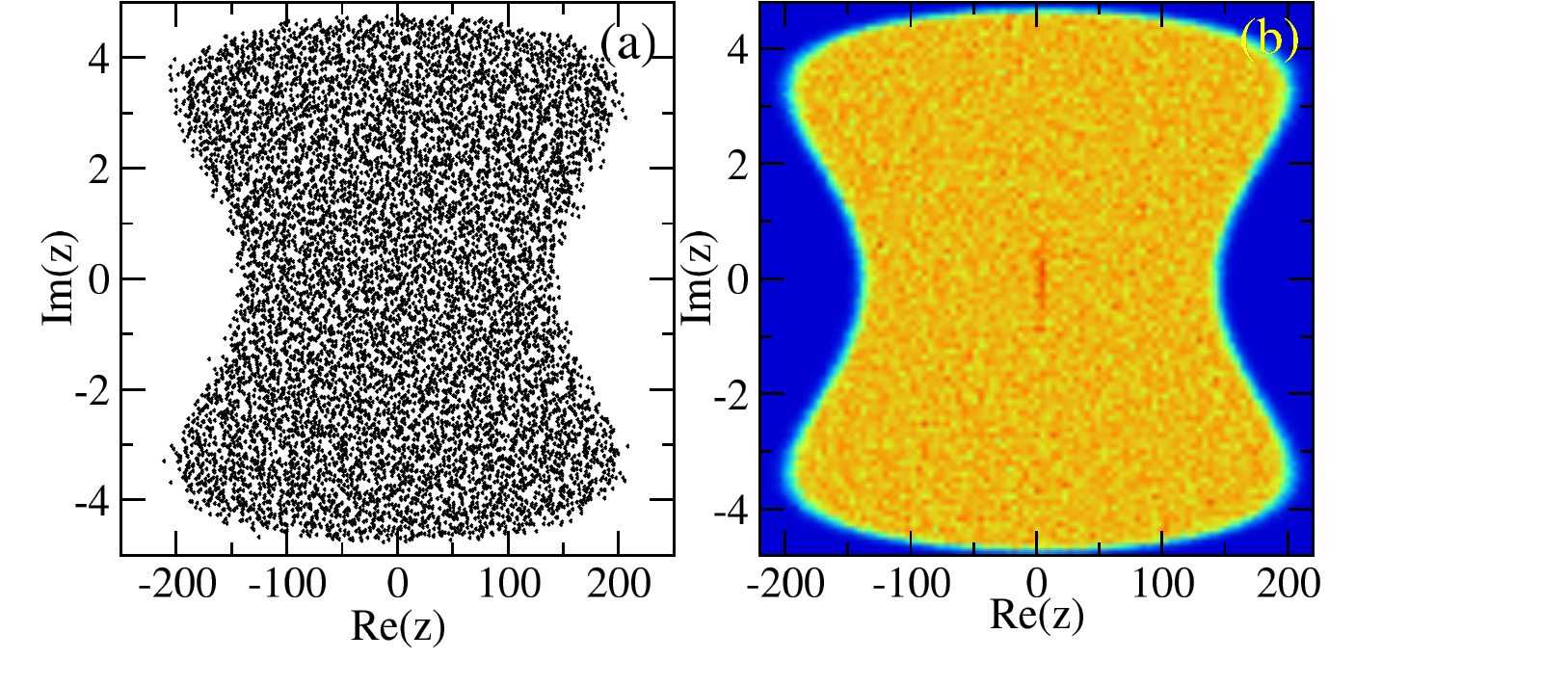}
	\caption{(a) The eigenvalue scatter plot, and (b) the spectral density for the spectra in Fig.~(\ref{fig-ev-r1}) unfolded using Eq.~(\ref{eq-unfolding-function-3a}). \label{fig-ev-r1-unfold-3}}
\end{figure}

We also test the unfolding discussed in ref.~\cite{Markum-1999} for weakly non-Hermitian systems. The jpd in Eq.~(\ref{eq-jpd}) correspond to a strongly non-Hermitian system because the eigenvalues have real and imaginary components of the same order. Rewriting Eq.~(5) of the main text in the Cartesian coordinates and separating in $x$ and $y$ variables, we get,
\begin{align}
\label{eq-unfolding-function-3}
\tilde dy & = dy,\\
~ \text{and}~\frac{1}{\pi}d\tilde x & = R_1(\sqrt{x^2+y^2}) dx.
\end{align}
We get after substituting Eq.~(\ref{eq-r1}) with $k = 2$ and integrating
\begin{align}
\label{eq-unfolding-function-3a}
\tilde y  = y
~\text{and}~\tilde x  =  \frac{4}{3} x^3 + 4 x y^2.
\end{align}
The unfolded eigenvalues and spectral density are shown in Fig.~(\ref{fig-ev-r1-unfold-3}). The average spectral density is constant everywhere except some spikes at the center. We avoid the central region to measure the number variance. As shown in Fig.~(\ref{fig-nvar-all}), it does not agree with number variance of the Ginibre ensemble.  

\bibliography{refs}